# What has been Revealed by Urban Grid Data of Shanghai

Yongkun Wang, Yaohui Jin, Bo Fan

China Institute for Urban Governance, Shanghai Jiao Tong University

ykw@sjtu.edu.cn, jinyh@sjtu.edu.cn, fanbo@sjtu.edu.cn

**Abstract**

*With the fast-growing economy in the past ten years, cities in China have experience great changes, meanwhile, huge volume of urban grid management data has been recorded. Studies on urban grid management are not common so far. This kind of study is important, however, because the urban grid data describes the individual behaviors and detailed problems in community, and reveals the dynamics of changing policies and social relations. In this article, we did a preliminary study on the urban grid management data of Shanghai, and investigated the key characteristics of the interactions between local government and citizen in such a fast-growing metropolitan. Our investigation illustrates the dynamics of coevolution between economy and living environments. We also developed mathematical model to quantitatively discover the spatial and temporal intra-relations among events found in data, providing insights to local government to fine tune the policy of resource allocation and give proper incentives to drive the coevolution to the optimal state, thereby achieving the good governance.*

# 1. INTRODUCTION

Shanghai, the biggest, richest and most cosmopolitan city of China, is positioned as a pioneer in the midst of the fastest and most intense process of urbanization the world has ever known (Greenspan, 2017). During this process, there are a lot of conflicts of interests and community problems among different sides (governments, companies, urban dwellers, etc.). To resolve conflicts and keep the economy growing in high speed, local governments and communities urgently need an efficient way.

To record and solve problems arising from local communities, Chinese government has deployed the Urban Grid Management System (UGMS) for more than ten years. UGMS divide the cities into many geographic grid cells, in which city dwellers can report issues found by hotlines. Therefore, governments can accurately locate the spot and solve the problem quickly. Local governments also train grid management workers to move around cells to collect the information with specially designed mobile devices. All these information is recorded in the databases. This IT-based UGMS system has greatly improved the efficiency of urban management. Chinese cities like Shanghai have experienced very intensive urbanization in the past ten years, with great and frequent changes in local communities, along with many conflicts of interests, which have been recorded by UGMS. With the development of big data technologies and open data campaigns, local governments are trying to open the government data including UGMS data, aim at insights from analysis of data, and seek to have a better way based on data to achieve effective and efficient governance.

The literature on government or urban data has grown substantially in recent years. There are a lot of research using urban data to provide solutions for smart city and smart governance, such as the research using air quality data (Clai 1998) (Zheng 2013), taxi data (Ferreira 2013) (Liu & Wang et. al. 2012), light data (Liu 2012), social network data (Simeone 2016), smart phone data (Calabrese 2013), etc. However, studies on UGMS data are not common. A few existing literatures on UGMS data has often focused on theory or policies (Jiang 2017) (Han 2017), instead of solid analysis on long-term real data. Existing literature fails to reveal the following important aspects which has significant influence on government policies: (1) the dynamics of coevolution between local government and communities in terms of development and living environment within certain time and space; (2) the intra relation of different problems are not well understood.

This article seeks to provide a study on large volume of UGMS data, which could supplement the existing literature with a new type of data. We seek to enhance our understanding of the coevolution of different interests of urban areas under the quick development of economy. Specifically, we try to find those aspects or interests that may have short-term conflicts during the development, and reveal the trend in time and space dimensions. To achieve this goal, we have created mathematical models and applied on large volume UGMS data to reveal the intra-relation of different problems and categories in both time and space dimension.

In general, our results suggest that the communities have the strong concern on environments, public order, and public transportation. The more developed the area is, the more concern arises. The spatial and temporal intra-relation analysis reveals the situations that may not be paid enough attention by local government. Although the local government hold the potential to push forward the local economy and foster a range of beneficial impacts on local communities, they might not get positive feedbacks from local communities if the demands and interests from the bottom of the society cannot be fully understood and satisfied. Therefore, in this article we identify those intensive interests and provide suggestions to government on how to make the best use of resources to produce results that meet needs.

The rest of this article will organize as follow: Section 2 will review related literature. Section 3 will provide our methodology and analysis, which is the core of this article. We will provide some findings and discussion in Section 4 and 5 respectively. Section 5 concludes this article.

## 2. LITERATURE REVIEW

There has been a lot of research literature on urban governance with classic study method, such as (Hoornbeek 2016) who did an in-depth analysis on local government collaboration with empirical study method. With the development of big data analysis technologies, researchers are trying to use urban data to get the findings in more details or bigger picture. By the data source, urban data research can be roughly divided into sensor-based (e.g., Internet of Things) or crowd-based (e.g., social media or citizen science) (Thakuriah 2017).

Sensor-based are mainly by the existing city stations which can generate data, such as meteorology stations, light stations, satellite data, public transportation data which can track the users' trajectories, etc. (Egami 2016) complemented and visualized the urban data for illegally parked bicycles. (Davoine 2014) showed us how to use their application to collect urban data about seismic risk. (Fauvel 2009) investigated using Kernel principal component analysis for feature extraction from hyperspectral remote sensing data. (Clai 1998) analyzed the atmospheric pollution data in the Bologna area with ARMA model. (Ren 2017) conducted spatiotemporal analyses of urban vegetation structural attributes using multi-temporal Landsat TM data and field measurements. (Zheng 2013) inferred air quality information based on the air quality data by monitor stations and other city data sources. (Ferreira 2013) explored the taxi trips data, and create model and system for users to explore and compare results. (Liu 2012) developed methods to analyze the Nighttime stable light data to discover the dynamics of urban expansion. (Taubenböck 2014) analyzed the satellite data to describe the mega-region area in a spatial method. (Liu & Wang 2012) revealed intra-urban land use variations from traffic patterns by analyzing taxi data.

With the quick development of crowd-sourcing and social networks, the characteristics of the city and the effects of the local policies can be well reflected by the responses from crowd-sourcing and social networks. (Wolff 2017) used games to motivate the citizen to engage with city data to help to design smart city solutions. (Schlesinger 2015) developed Urban–Rural Index (URI) by using the crowd-sourced data on satellite images and OpenStreetMap data, which helps to understand urban development patterns. (Simeone 2016) shared experience on Urban Sensing research project, showing the potential and limitations of social network analysis and data visualization as research methods in urban studies. (Kanhere 2013) analyzed the major research challenges by using mobile phone data generated in urban spaces. (Calabrese 2013) analyzed the mobile phone data for transportation research, to understand the intra-urban variation of mobility. (Rathore 2016) proposed a system to deploy sensors and analyze collected data aiming for smart city.

There is no clear border between the sensor-based and crowd-based data. We also see some literatures using both sensor-based and crowd-based data. (Panagiotou 2016) had a good practice in the city of Dublin to explore how to do urban data management effectively, by using traffic data and social network data (Twitter).

Unlike the above sensor-based or crowd-based data, urban grid data is a special kind of data which contains both sensor data actively collected by trained staffs, and a lot of crowd-based data by dwellers via hotlines. There has been little literature on Urban Grid data analysis, most of the existing research is focusing the theory and system. (Jiang 2017) studied the conversion of "Urban Grid Management" to "Urban Grid Governance", including conversion from "management" to "service", from "maintaining stable" to "satisfying demands", from "single direction management" to "double direction communication", etc. (Han 2017) studied how to clear the urban governance tasks, processes by using urban grid management, and limitations for this method. (Ni 2011) introduced the architecture and technologies of urban grid management system used in Beijing.

This article tries to complement the existing literature by providing the in-depth analysis on urban grid data.

## 3. METHODOLOGY AND ANALYSIS

In this section, we provide the basic statistics on the UGMS data we got, as well as the spatial and temporal analysis to find the macro behaviors/interests of civilians. We also design a mathematical model to perform a 'self-comparative' analysis, to find the evolution of events or states.

### 3.1 Urban Grid Management System

Urban Grid Management System (UGMS) has been deployed in China for more than ten years. UGMS uses GIS, GPS, and mobile information technology to divide the cities into many geographic grid cells, with each cell is about 100 by 100 meters. Combined with the geographic information system, each cell was given a unique code, and public properties within the cell, such as buildings, lampposts, and etc., were catalogued, coded, located and saved in databases (Wu, 2014). Civilians in each cell can report their complains or findings to the UGMS offices by hotlines, and officers will urge the corresponding departments of local governments to solve the problem. Once the problem is solved, the UGMS office staff will call back to the reporter and inform him/her the result. Upon receiving the result, the reporter can close the issue if he/she is happy with it and give good rating about the service this time, or bring up new requirements to the office staff and start the new cycle of problem-solving process. UGMS office also trains the grid management workers to move around cells to collect the information with specially designed mobile devices, which can find issues proactively before being complained by the local dwellers.

All our analysis is drawn from a dataset with a time span of 20 months (2015-01-01 to 2016-08-31), total number of cleaned records is 1,131,423. The data came from a district of Shanghai, with more than 5 million residents. This district contains various types of areas, such as a downtown area with some of the best-known buildings, high-tech area with offices of some best-known high-tech and internet companies, resort area, port area, nature reserve area, etc. Therefore, the dataset covers almost all types of community complain/conflict cases of Shanghai. Each record of the data is composed of tens of fields including the *urban grid cell id, timestamp, starting time, end time, longitude and latitude, reporting types (by hotlines or mobile devices of workers), address, description, category, priority, feedback*, etc. The *starting time* tells us when the problem is found, and the *end time* shows us when the problem is solved.

## 3.2 Basic Statistics about Dataset

Table 1 provides the statistics of the contents in the dataset. We can see that the major part (79%) of the data is from the "Mobile Device" of workers trained by local governments. This shows that the local governments are very proactive to identify and prevent the potential problems. Most of the problems reported from "Mobile Device" fall in the following two categories: 42% of problems fall in the category of "Environment, City Appearance", which includes findings related to illegal ads, uncleaned garbage, illegal shops, violation of airing cloths/bedding, construction violation, etc.; 24% of problems fall in the category of "Street Order", which includes the findings related to illegal selling, construction, and parking on streets.

The complains by "Hotline" is about 21% of total records. People mainly complain in " Greening, City Appearance, Construction, Housing, Environment, Rivers, Public Transportation", and some people reported some problems related to " Management of human resource related affairs ".

These basic statistics clearly show that the living environmental conflicts are the major part of problems coming along with the quick development of local economy. The environmental problems co-evolve with the development of local economy, so the economy development must adapt to the environment. Otherwise, we would see in later analysis that more and more conflicts and complains related to local environments would be brought up to the government. To achieve good urban governance, the government must achieve a quality of life sought by the residents, which is mainly the quality of living environment.

Table 1, Description of Dataset

| Source | Percentage % | Category | Percentage % |
|---|---|---|---|
| Mobile Device | 79 | Environment, City Appearance | 42 |
| | | Street Order | 24 |
| | | Public Facility | 4 |
| | | Greening | 2 |
| | | Road, Traffic | 2 |
| | | Resident Area Mgmt. | 1 |
| | | Others | 4 |
| Hotline | 21 | Greening, City Appearance, Construction, Housing, Environment, Rivers, Public Transportation | 17 |
| | | Management of human resource related affairs | 1 |
| | | Others | 4 |

## 3.3 Temporal and Spatial Analysis

In order to find the trend in different categories, we sum up the data in each week and plot them in Figure 1 to find the temporal trend of different types of problems or events. The x-axis shows the time in week which spans almost two years (0-52 is year 2015 and 52-90 is year 2016). The y-axis is the number of events or problems. Each sub-figure shows the histogram of a specific category as described by the sub-figure title. We can see that the number of events in all categories varies on a seasonal base, that is, the number of events reaches peak point around summer (week 20-40), and slumps in winter (week 50-60). This is consistent with the seasonal variations in physical activity of human (Shephard and Aoyagi 2009). We can take the "Greening" category as an example to explain the reason of seasonal variations. The "Greening" category shows that people have more requirements/complains around summer than that in winter which can be explained by more activities (either economic or physical) around summer in greening areas like gardens. Therefore, the seasonal variations of human activities are one dynamic driving the developments and conflicts. It would be more efficient for the short-term incentives or policies to consider the seasonal factors to reduce the number of problems or complains.

Figure 1 also shows that the number of problems are almost doubled after week 60 (year 2016) in the following categories such as "Pub Facility", "Road, Traffic", "Environment", "Street Order". We talked with the office staffs and confirmed that they launched a new IT system which greatly improved the efficiency of collecting the problems.

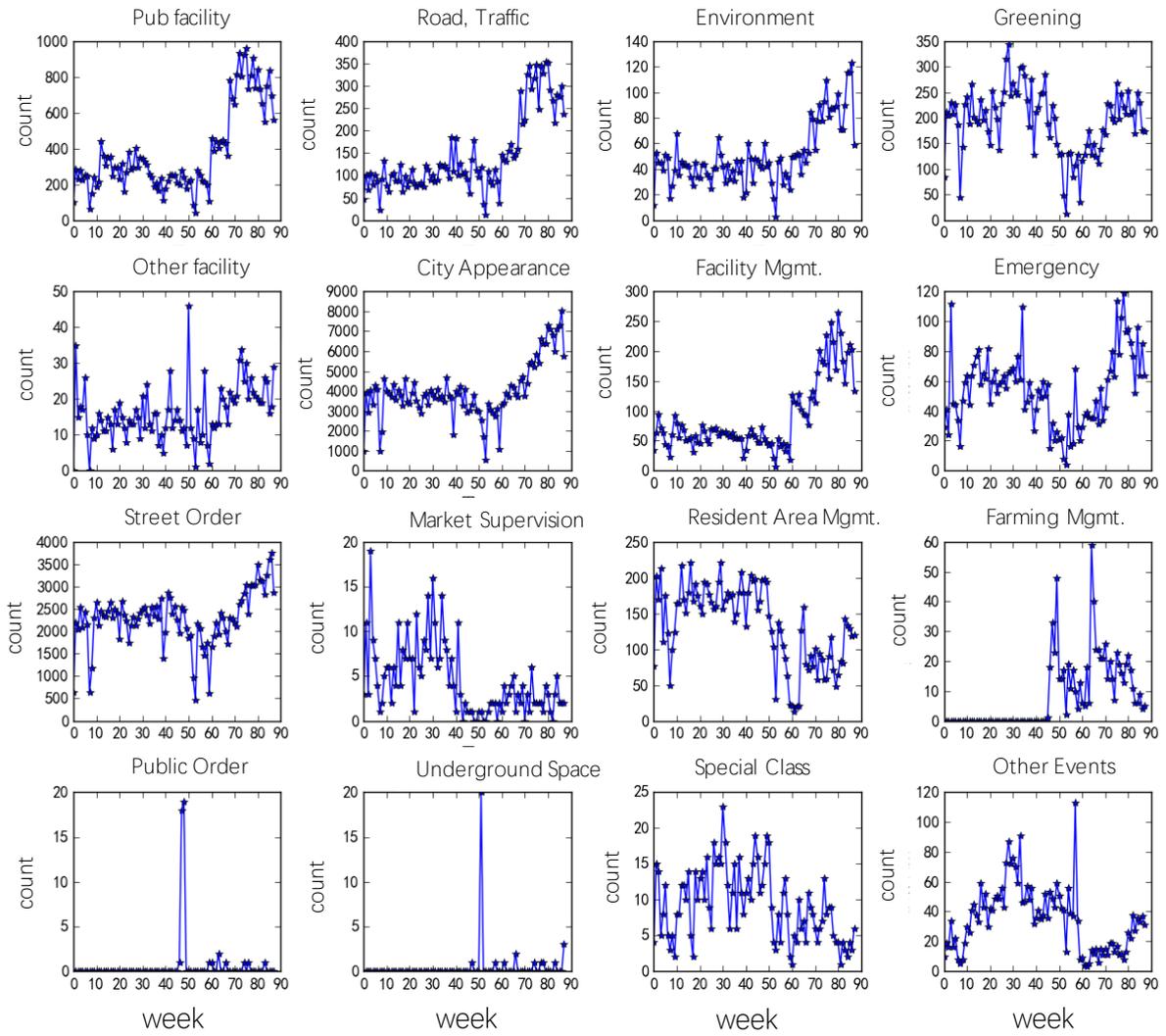

Figure 1 Trend of the number of reported problems counted by week. x axis is in week.

We also have interests to know the geographic distribution of the problems, which may help the local government to optimize the resource allocation by locations. Figure 2 shows the normalized spatial distribution of the reported problems. The x-axis and y-axis represent normalized longitude and latitude respectively, and the sub-figure title shows the category of problems. The darker the color is, the more events in that area. We find that the number of problems is proportional to the population and intensity of economic activities of the area. The dark color area is the center of downtown area in most of the categories except the 'Farming Mgmt.' category, which can be easily understood as the farms are usually far away from the center of downtown. The 'Resident Area Mgmt.' category clearly shows that both the downtown and suburb area have many resident related problems, which is consistent with the distribution of residential areas. This figure clear shows that the problems are tightly co-evolving with the local economy; the higher the level of economic development, the more problems reported. This is not desirable for local government. The byproduct of high-speed development is the damage to the living environments. Balanced resolution for the conflicts between economy and ecosystem should be placed with higher priority when policies are made, to develop both economy and environment harmoniously to an optimal state. Mediating differing interests in economy and environment to reach a broad consensus, would help local government to achieve good governance.

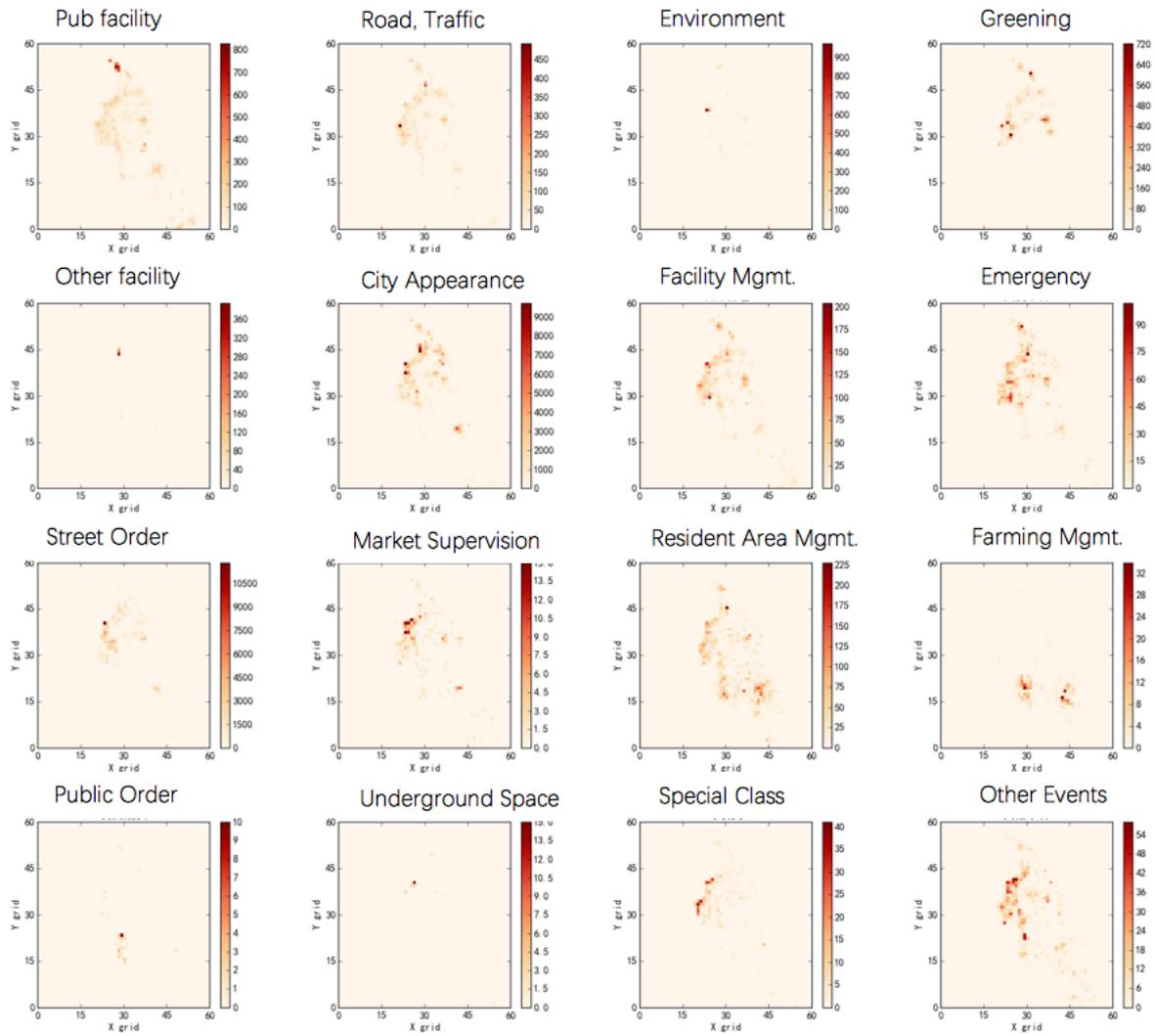

Figure 2 Normalized Spatial Distribution of Problems

## 3.4 Self-Comparative Analysis

To get statistical comparative analysis (e.g., compare different regions over time) and study the evolution of problems, we create the temporal and spatial relevance models to describe the relevance of states in different time and areas.

Temporal relevance of events is defined as the relation between the state of time $t_i$ and state of time $t_j$, where the state is the event distribution on an area. For the formal mathematical definition, please check the ENDNOTES section. Temporal relevance model can show the evolution of overall state of different types of events, including those related to economy and living environments. Figure 3 shows the event state as time being elapsed. The x-axis is time in day, and y-axis is the relevance index. The higher score of index, the more relevant. We can see that at the time close to the initial state, the relevance score is high; as time elapse, the relevance score is decreasing. This shows that the problems are solved gradually, so that later state is less relevant to the initial state. We also can see that some lines are not decreasing smoothly, but have many peaks periodically, which means some problems have been brought up again and again. This indicates that some problems cannot be solved in one shot, but solved gradually with solutions adjusted by feedbacks from reporters. Especially the problems in category of "Environment", "Greening", "Other facility", "City Appearance", the relevance score is back to the initial state from time to time, which indicates that the later problems are almost identical to the initial ones and the reporters raise their problems again if there are no acceptable solutions. It suggests that the local government need to find the root cause of these types of problems and reduce the number of cycles before getting the final solution, otherwise, people may not be happy with the responsiveness and accountability of the government. Some policies may also be needed to give both government divisions and local residents incentives to solve these problems together. This will activate and broaden the participation of local residents, and help them participate constructively, which is an importance characteristic of the good governance.

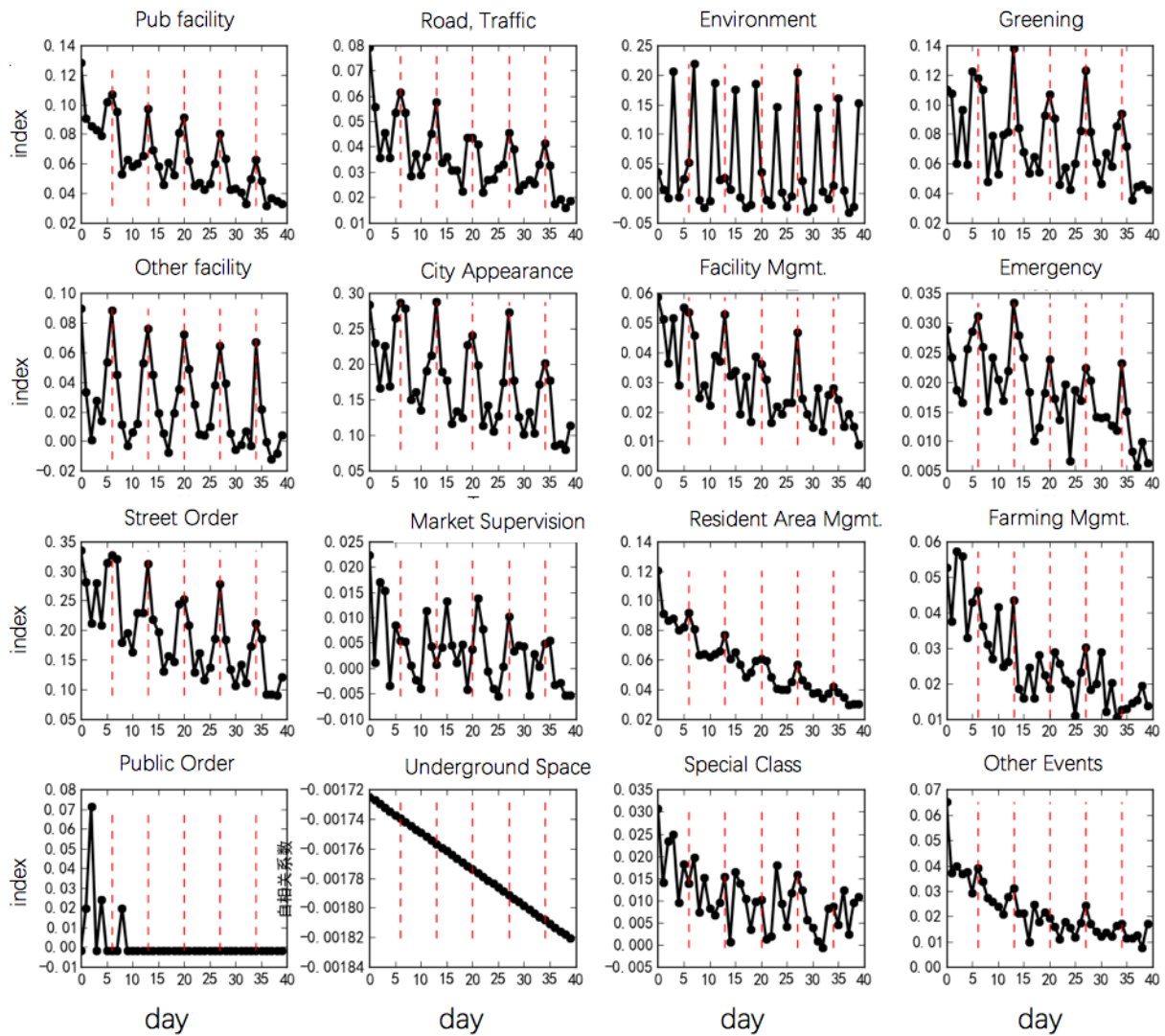

Figure 3 Temporal relevance of each category

Now we study the state changes with respect to the locations. We define the spatial relevance of events as the similarities between event state in area $s_i$ and event state in area $s_j$. The formal mathematical definition can be found in ENDNOTES section. The spatial relevance shows the event state varies across different areas. Figure 4 shows the spatial relevance score of event state within certain distance. The x-axis is the distance measured by 10km (1e4, 10,000 meters), y-axis is the spatial relevance index. The higher score of index, the more relevant. Overall, we can see that most of lines are decreasing sharply, which means that the influence of the problems is usually limited to certain spots, instead of broadcasting to a wide area. However, we also can see that the lines of the following categories, "Greening", "City Appearance", "Facility Mgmt.", "Emergency", "Street Order", etc., decrease much slower than that of others, which implies that these categories of problems would also happen in nearby areas. It provides a hint to perform early warning for these types of problems if we found similar problems in nearby area. It is interesting to see that there is a curve peak for "Resident Area Mgmt." around 5 (10km), which implies that the problems would happen with high possibility in another resident area. This is consistent with the spatial distribution in Figure 2, in which it clearly shows there are several dense residential areas in downtown and suburb. It tells us that the problems and conflicts, which is the by-product of quick development, are more prominent in areas with dense population, which is also a strong driving force to provide the perfect complement to the development.

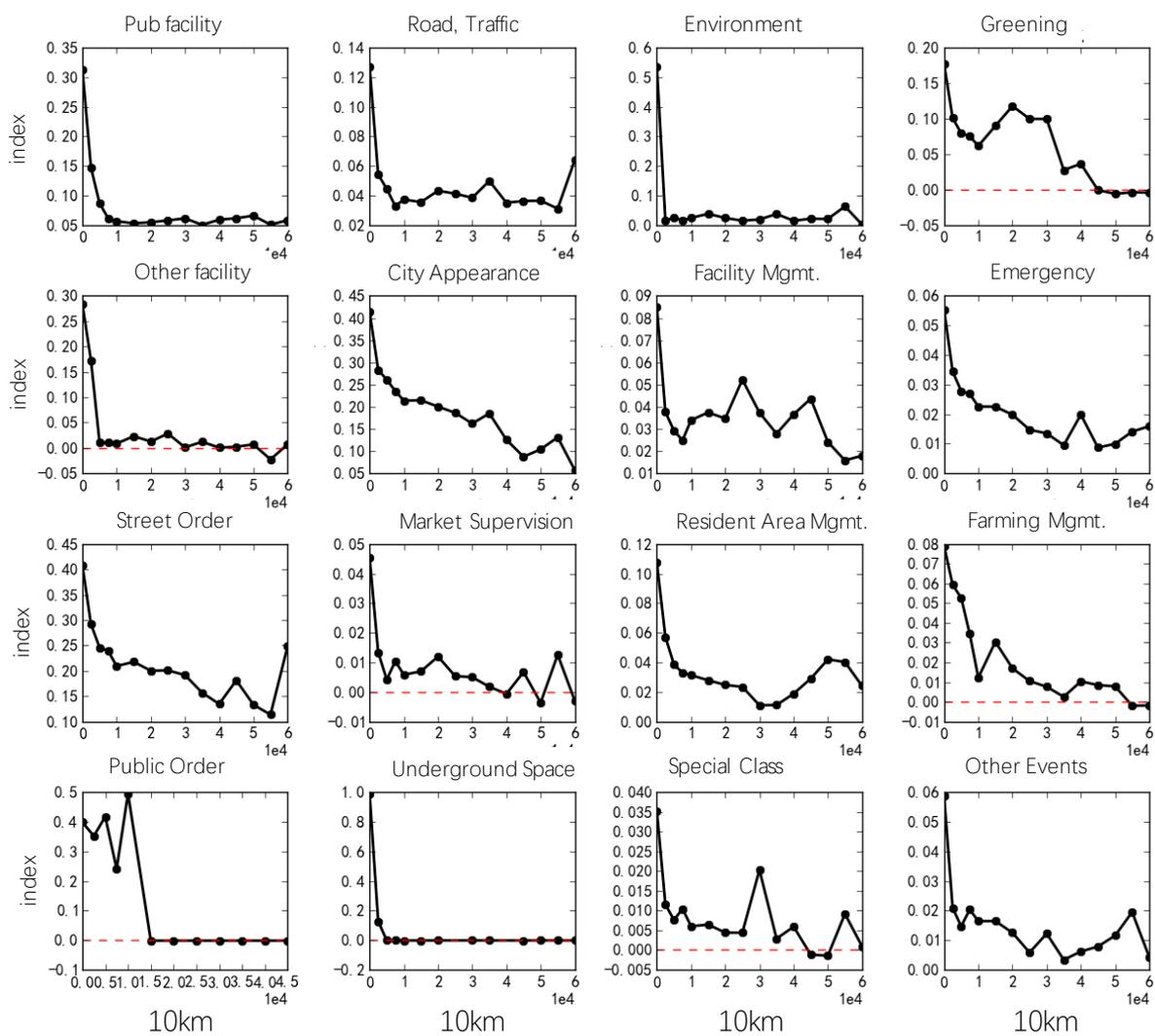

Figure 4 Spatial relevance within each category

## 3.5 Category Relevance Analysis

All the events or problems reported to UGMS office will be classified into some categories or classes. The categories are pre-defined by specialist when the UGMS system was launched. Over time, there are more and more new categories added along with the development. There are also some categories whose scope might be changed. So the UGMS officers might need an efficient way to cluster and clean the categories. What's more, we have great interest to know the relevance of different categories, since many of those events or problems might not happen alone.

We quantify the category relevance by calculating the mutual information of each category. Specifically, we calculate the entropy of each category with the probabilistic distribution across cells, then calculate the mutual information of every category pair. The detailed mathematical definition can be found in ENDNOTES section. The purpose of category relevance is not only cluster the categories, but also find some interesting relations between categories which may usually be considered as irrelevant. Figure 5 shows the category relevance results from both Mobile Devices by workers and Hotlines by citizens. The darker cell color means more relevant between two categories. We can see some strong relations marked in area A, B and C. As shown in area A, the 'Underground' categories by 'Mobile Devices' have strong relation with 'Police' and 'Construction' classes of 'Hotlines'. The former pair, 'Underground - Police', can be explained that the changes to underground structures or facilities may cause people to complain on hotlines as a police (firefighting) issues. It might be easy to understand the latter one, 'Underground - Construction', as the events found by workers about the damage to underground facilities or structures, might be reported by residents as a construction problems. As seen in area B, the 'Underground' issues are also related to 'Well Cover' related issues, as both of them are underground. C area shows that 'Greening' problems (such as destroying garden, public green land, etc.) is connected with the 'Construction' by Hotlines. This helps us to find the real reason when local residents complained to hotlines without knowing the cause. The 'Construction' and 'Police' class are related closely with 'Underground (including civil defense facilities)' and 'Greening (include resident, company, and public grassy area)' categories, which reflects the conflicts of development in economy and environment.

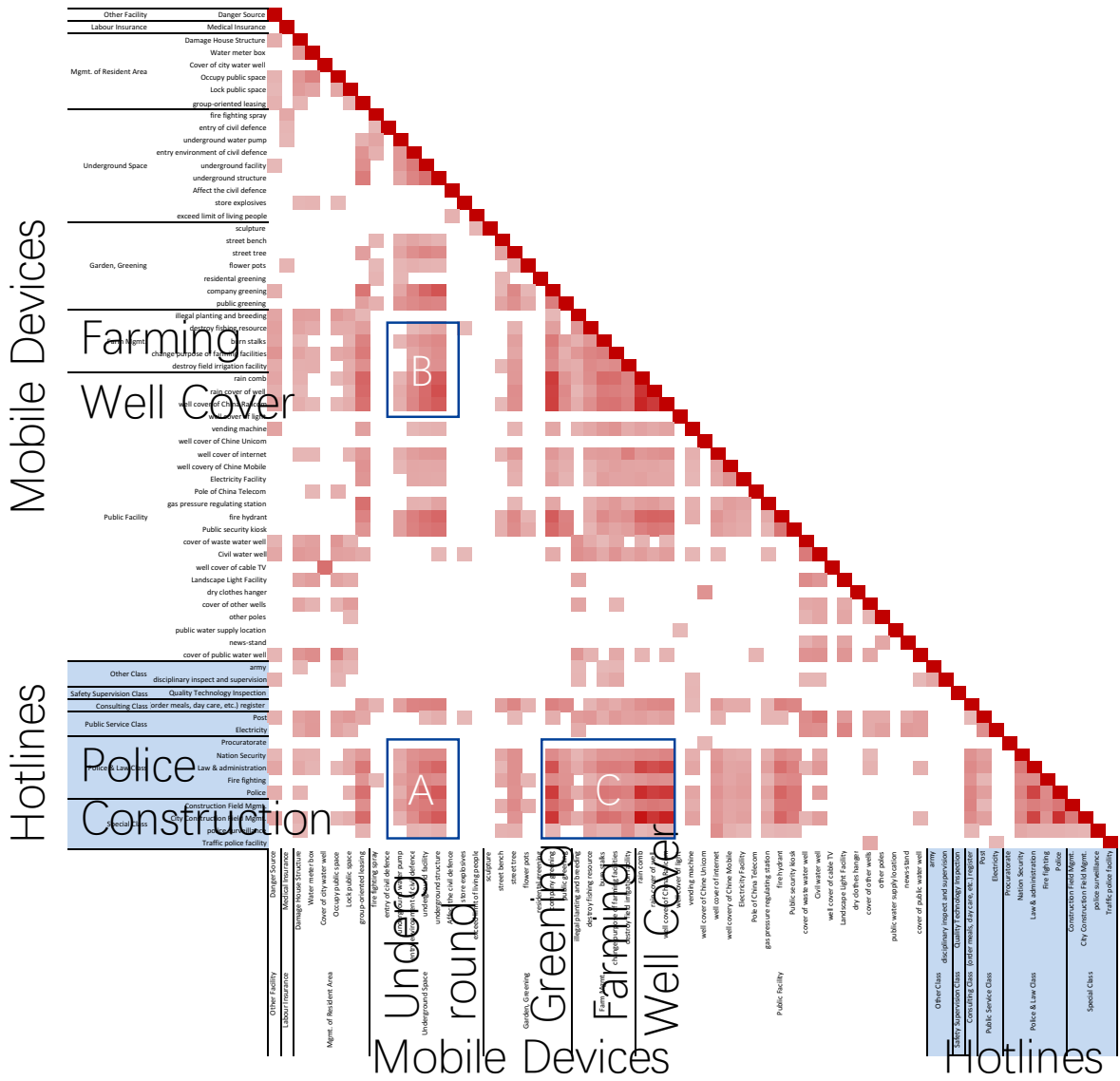

Figure 5 Category Relevance Analysis

## FINDINGS

The methodology and analysis presented above show the successes and conflicts among different sides during the development of local communities. Besides the analysis, we have the following findings.

### VAGUE SCOPE

Table 1 shows that the complains or problems are mainly in a few categories, which shows that local residents have similar concern on certain problems. But this also tells us that the scope of categories may not be well defined, thereby each category covers too many problems. This is serious since large category implies vague scope, which might be difficult for stuff to quickly locate the corresponding government divisions to solve the problem.

The problem of vague scope of categories is also brought up by Figure 3, which tells us that some problems were brought up again and again during the solving process. After talking with the staffs in UGMS, we found that there are many cases are returned by divisions who thought that those cases were out of the scope of their responsibilities.

In order to improve the efficiency of these categories, we recommend to fine tune the scope of categories, create more sub-categories with detailed scope. This can be done by learning from the data. Clustering the data and filter the sub-categories by machines firstly, then the specialist can fine tune and optimize the generated sub-categories. Once the machine learns the features of these sub-categories, it can help to classify the problems automatically, and link the suggested divisions quickly to solve the problem.

### PREDICTION

The temporal and spatial analysis in Figure 2 and Figure 3 shows that the number of events or problems is not evenly distributed across time and space. For those days and locations with more events, we suggest to allocate more resources in advance to be prepared. An early warning system can help to predict the time and location for local government to do resource allocation. This can be done by learning the features of data series to create a model for prediction.

### ADJUSTMENT

In category relevance analysis section, Figure 5 shows that some categories are closely connected. This shows that, on the contrary to Finding 1, some categories overlap each other, and can be cleaned, merged or adjusted. With less categories, local government can gather their limited resources and re-organize their hands to pay more attention on urgent affairs.

In summary, these findings show that there are still a lot of efforts required to fine tune the UGMS and related policies.

## 4. DISCUSSION

Although we try to analyze the UGMS data by many state-of-the-art techniques, and talk with staffs to get as more information as possible beside the data from their databases, there are still some questions which are hard to be answered. For example, the progress of solving some problems are very slow so that the UGMS staffs or citizens have to urge the corresponding divisions. It remains unknown to us that why some divisions are slow on solving these cases. Our conjecture is that the economy is still the key performance indicator of some divisions, so that they might try to find a balanced solution to avoid the conflicts when solving the problems which might slow down the development. The answer to these kinds of questions cannot be found by analyzing the urban data. We might need to use the classic research method to delicately design a comparative study on two or more experimental regions to find the answer.

## CONCLUSION

We provide a study on urban data which has not been well studied by existing literatures. In spite of the limitations of data analysis, this study provides a big picture on the urban grid management system in one of the largest metropolitan. Drawing on the full data from the latest 20 months of an area with various types of regions, we demonstrate the temporal and spatial characteristics of local communities within the urban grid, showing the co-evolution of economy, environments, and other entities that remain conflicted with each other. To get a more in-depth analysis, we developed temporal and spatial relevance model for urban grid data, which further reveals some special characteristics like periodicity and locality. We also define a quantitative index to evaluate the relations between different categories, which is very helpful for local government to not only discover the potential relations between different events or problems, but also clean the legacy data.

Our findings can help local governments to allocate their limited resource in advance to some problems in advance to prevent from getting worse. The policymakers can also identify ways to develop certain incentives to those divisions whose may hesitate to solve some problems which might decrease the rate of economic growth, which eventually will lead to the optimal state of economy and other conflict aspects like environment. With these suggestions, it could help to approximate to good governance by increased responsiveness, accountability of local government.

Our study on data also help the local governments to fully understand the benefits of technologies which can help to clean huge amount of legacy data, automatically classify and dispatch the tasks to the most related divisions. The analysis on data also shows that we can generate early warnings before the problems turning into emergencies. We believe that the follow-up research effort on these directions can greatly improve the efficiency of local government and the urban grid management system.

# ENDNOTES

1. Spatial and temporal relevance model

Assume spatial and temporal distribution of the data is $\{Y(s,t): s \in D_s, t \in D_t\}$, where $s, t$ represent the spatial and temporal variable respectively. The time series in space $s_i$ and $s_j$ is $Y(s_i,t)$、$Y(s_j,t)$, therefore, the temporal self-relevance function of area $s_i$ can be defined as

$$C(s_i,\tau) = \left(Y(s_i,t) - \mu(s_i)\right)\left(Y(s_i,t-\tau) - \mu(s_i)\right)$$

In which the global temporal self-relevance function can be defined as

$$C(\tau) = \frac{\sum_{s_i \in D_s} |\mu(s_i)| \left(Y(s_i,t) - \mu(s_i)\right)\left(Y(s_i,t-\tau) - \mu(s_i)\right)}{\sum_{s_i \in D_s} \mu(s_i)}$$

Temporal series self-relevance function of area $s_i$, $s_i$ can be defined as:

$$C(s_i,s_j) = \left(Y(s_i,t) - \mu(s_i)\right)\left(Y(s_j,t) - \mu(s_j)\right)$$

In which the global spatial mutual relevance function can be defined as

$$C(d_k) = \frac{\sum_{s_i \in D_s, s_j \in D_s, d(s_i,s_j) = d_k} |\mu(s_i)||\mu(s_j)|\left(Y(s_i,t) - \mu(s_i)\right)\left(Y(s_j,t) - \mu(s_j)\right)}{\sum_{s_i \in D_s, s_j \in D_s, d(s_i,s_j) = d_k} |\mu(s_i)||\mu(s_j)|}$$

2. Category Mutual Information Model

Assume two category vectors, $U$ and $V$. Their entropy is the amount of uncertainty for a partition set, defined by:

$$H(U) = -\sum_{i=1}^{|U|} P(i)\log(P(i))$$

where $P(i) = |U_i|/N$ is the probability that an object picked at random from $U$ falls into class $U_i$, here it is the category distribution over the urban grid cells. Likewise for $V$:

$$H(V) = -\sum_{j=1}^{|V|} P'(j)\log(P'(j))$$

With $P'(j) = |V_j|/N$. The mutual information (MI) between $U$ and $V$ is calculated by:

$$\text{MI}(U,V) = \sum_{i=1}^{|U|}\sum_{j=1}^{|V|} P(i,j)\log\left(\frac{P(i,j)}{P(i)P'(j)}\right)$$

where $P(i,j) = |U_i \cap V_j|/N$ is the probability that an object picked at random falls into both classes $U_i$ and $V_j$, here it is number of events in these two categories fall in the same cell.

It also can be expressed in set cardinality formulation:

$$\text{MI}(U,V) = \sum_{i=1}^{|U|}\sum_{j=1}^{|V|} \frac{|U_i \cap V_j|}{N} \log\left(\frac{N|U_i \cap V_j|}{|U_i||V_j|}\right)$$

The normalized mutual information is defined as

$$\text{NMI}(U,V) = \frac{\text{MI}(U,V)}{\sqrt{H(U)H(V)}}$$

# REFERENCES

Greenspan, A. (2014). Shanghai Future: Modernity Remade. (1st ed.). Oxford University Press

Wu, Q. (2014). Urban Grid Management and Police State in China: A Brief Overview. Retrieved from https://chinachange.org/2013/08/08/the-urban-grid-management-and-police-state-in-china-a-brief-overview/

Hoornbeek, J., Beechey, T. and Pascarella, T. (2016), fostering local government collaboration: an empirical analysis of case studies in ohio. journal of urban affairs, 38: 252-279

Thakuriah P.., Tilahun N.Y., Zellner M. (2017) Big Data and Urban Informatics: Innovations and Challenges to Urban Planning and Knowledge Discovery. In: Thakuriah P., Tilahun N., Zellner M. (eds) Seeing Cities Through Big Data. Springer Geography. Springer, Cham

Egami S., Kawamura T., Sei Y., Tahara Y., Ohsuga A. (2016) A Solution to Visualize Open Urban Data for Illegally Parked Bicycles. In: Hameurlain A. et al. (eds) Transactions on Large-Scale Data- and Knowledge-Centered Systems XXVII. Lecture Notes in Computer Science, vol 9860. Springer, Berlin, Heidelberg

Davoine PA., Gensel J., Gueguen P., Poulenard L. (2014) Isibat: A Web and Wireless Application for Collecting Urban Data about Seismic Risk. In: Pfoser D., Li KJ. (eds) Web and Wireless Geographical Information Systems. W2GIS 2014

Fauvel, M., Chanussot, J. & Benediktsson, (2009) Kernel Principal Component Analysis for the Classification of Hyperspectral Remote Sensing Data over Urban Areas. J.A. EURASIP J. Adv. Signal Process. (2009) 2009: 783194

Clai, G., Kerschbaumer, A., Tosi, E. et al. (1998) Analysis of Urban Atmospheric Pollution Data in the Bologna Area. Environ Monit Assess (1998) 52: 149.

Schlesinger J. (2015) Using Crowd-Sourced Data to Quantify the Complex Urban Fabric—OpenStreetMap and the Urban–Rural Index. In: Jokar Arsanjani J., Zipf A., Mooney P., Helbich M. (eds) OpenStreetMap in GIScience. Lecture Notes in Geoinformation and Cartography. Springer, Cham

Ren, Z., Pu, R., Zheng, H. et al. (2017) Spatiotemporal analyses of urban vegetation structural attributes using multitemporal Landsat TM data and field measurements. Annals of Forest Science (2017) 74: 54.

Yu Zheng, Furui Liu, and Hsun-Ping Hsieh. (2013). U-Air: when urban air quality inference meets big data. In Proceedings of the 19th ACM SIGKDD international conference on Knowledge discovery and data mining (KDD '13)

Nivan Ferreira, Jorge Poco, Huy T. Vo, Juliana Freire, and Cláudio T. Silva. 2013. Visual Exploration of Big Spatio-Temporal Urban Data: A Study of New York City Taxi Trips. IEEE Transactions on Visualization and Computer Graphics 19, 12 (December 2013), 2149-2158.

Zhifeng Liu, Chunyang He, Qiaofeng Zhang, Qingxu Huang, Yang Yang, (2012) Extracting the dynamics of urban expansion in China using DMSP-OLS nighttime light data from 1992 to 2008, In Landscape and Urban Planning, Volume 106, Issue 1, 2012, Pages 62-72

M. Mazhar Rathore, Awais Ahmad, Anand Paul, and Seungmin Rho. 2016. Urban planning and building smart cities based on the Internet of Things using Big Data analytics. Comput. Netw. 101, C (June 2016), 63-80

H. Taubenböck, M. Wiesner, A. Felbier, M. Marconcini, T. Esch, S. Dech, (2014) New dimensions of urban landscapes: The spatio-temporal evolution from a polynuclei area to a mega-region based on remote sensing data, In Applied Geography, Volume 47, 2014, Pages 137-153

Yu Liu, Fahui Wang, Yu Xiao, Song Gao, (2012) Urban land uses and traffic 'source-sink areas': Evidence from GPS-enabled taxi data in Shanghai, In Landscape and Urban Planning, Volume 106, Issue 1, 2012, Pages 73-87

Wolff A. et al. (2017) Engaging with the Smart City Through Urban Data Games. In: Nijholt A. (eds) Playable Cities. Gaming Media and Social Effects. Springer, Singapore

Simeone L., Patelli P. (2016) Urban Sensing: Potential and Limitations of Social Network Analysis and Data Visualization as Research Methods in Urban Studies. In: Kubitschko S., Kaun A. (eds) Innovative Methods in Media and Communication Research. Palgrave Macmillan, Cham

Kanhere S.S. (2013) Participatory Sensing: Crowdsourcing Data from Mobile Smartphones in Urban Spaces. In: Hota C., Srimani P.K. (eds) Distributed Computing and Internet Technology. ICDCIT 2013.

Francesco Calabrese, Mi Diao, Giusy Di Lorenzo, Joseph Ferreira, Carlo Ratti, (2013) Understanding individual mobility patterns from urban sensing data: A mobile phone trace example, In Transportation Research Part C: Emerging Technologies, Volume 26, 2013, Pages 301-313

Panagiotou N. et al. (2016) Intelligent Urban Data Monitoring for Smart Cities. In: Berendt B. et al. (eds) Machine Learning and Knowledge Discovery in Databases. ECML PKDD 2016

X. Jiang, Y. Jiao (2017) From "Urban Grid Management" to "Urban Grid Governance", Theoretical Investigation, 2015,(06):139-143 (in Chinese)

Z. Han (2017) Clearance and Limitation of Urban Governance: Analysis by Urban Grid Management, Exploration and Free Views, 2017,(09):100-107 (in Chinese)

D. Ni, (2011) Research on Urban Grid Management Platform based on Cloud Service for Social Service, E-Government, 2011,(10):86-91 (in Chinese)

Shephard, R.J. & Aoyagi, Y. (2009). Seasonal variations in physical activity and implications for human health. Eur J Appl Physiol (2009) 107: 251.